\documentclass{revtex4}
\usepackage{amsmath,amsthm}
\usepackage{rotating}
\usepackage{multirow}
\usepackage{graphicx}

\usepackage{amsfonts}
\newtheorem{theorem}{Theorem}[section]
\newtheorem{lemma}[theorem]{Lemma}
\newtheorem{proposition}[theorem]{Proposition}

\theoremstyle{remark}

\theoremstyle{definition}
\newtheorem{definition}[theorem]{Definition}

\theoremstyle{example}
\newtheorem{example}[theorem]{Example}

\theoremstyle{notation}

\newcommand{\bra}[1]{\langle#1|}
\newcommand{\ket}[1]{|#1\rangle}

\begin{document}

\title{ Rescaling transformations and the Grothendieck bound formalism in a single quantum system}      
\author{A. Vourdas}
\affiliation{Department of Computer Science,\\
University of Bradford, \\
Bradford BD7 1DP, United Kingdom\\a.vourdas@bradford.ac.uk}

\begin{abstract}
The Grothedieck bound formalism is studied using `rescaling transformations', in the context of a single quantum system.
The rescaling transformations enlarge the set of unitary transformations (which apply to isolated systems), with transformations that change not only the phase but also  the absolute value of the wavefunction,
and can be linked to irreversible phenomena (e.g., quantum tunnelling, damping and amplification, etc).
A special case of rescaling transformations are the dequantisation transformations, which map a Hilbert space formalism into a formalism of scalars.
The Grothendieck formalism considers a `classical' quadratic form ${\cal C}(\theta)$ which takes values less than $1$, and the corresponding `quantum' quadratic form ${\cal Q}(\theta)$
which takes values greater than $1$, up to the complex Grothendieck constant $k_G$.
It is shown that ${\cal Q}(\theta)$ can be expressed as the trace of the product of $\theta$ with two rescaling matrices, and ${\cal C}(\theta)$ can be expressed as
the trace of the product of $\theta$ with two dequantisation matrices. 
Values of ${\cal Q}(\theta)$ in the `ultra-quantum' region  $(1,k_G)$ are very important, because this region is classically forbidden (${\cal C}(\theta)$ cannot take values in it). 
An example with ${\cal Q}(\theta)\in (1,k_G)$ is given,
which is related to phenomena where classically isolated by high potentials regions of space, communicate through quantum tunnelling.
Other examples show that `ultra-quantumness' according to the Grothendieck formalism (${\cal Q}(\theta)\in (1,k_G)$), 
is different from quantumness according to other criteria (like quantum interference or the uncertainty principle).

\end{abstract}

\maketitle

\section{Introduction}

An important inequality in pure mathematics is the Grothendieck inequality
which originally was formulated \cite{GR1} in the context of a tensor product of Banach spaces.
For this reason applications in quantum physics\cite{E1,E2,E3,E4,E5,E6,E7,E8}  have been in the context of entangled multipartite systems described by tensor products of Hilbert spaces.
Later mathematical work \cite{GR2,GR3,GR4,GR5,GR6,GR7} emphasised that the Grothendieck theorem can be formulated outside tensor product theory.
This motivated our work in refs\cite{VV,VV1} that uses the Grothendieck bound formalism in a single quantum system, and is unrelated to multipartite entangled systems.

The Grothendieck formalism considers a `classical' quadratic form ${\cal C}$ that uses complex numbers in the unit disc (scalars).
It also considers a `quantum' quadratic form ${\cal Q}$   that replaces the scalars with vectors in the unit ball of a Hilbert space.
The Grothendieck theorem shows that if the classical ${\cal C}\le 1$, the corresponding quantum ${\cal Q}$ might take values greater than $1$, up to the complex Grothendieck constant $k_G$.
The  region  $(1,k_G)$ is `ultra-quantum' because  ${\cal Q}$ might take values in it, while ${\cal C}$ cannot  take values in it (it is classically forbidden region). 

Many important quantum phenomena give ${\cal Q}<1$, e.g., oscillatory behaviour related to quantum interference, or quantum phenomena expressed through uncertainty relations (squeezing). 
The Grothendieck `ultra-quantumness' (${\cal Q}\in (1,k_G)$)
is different from quantumness in isolated systems, such as quantum interference or quantumness expressed through uncertainty relations
(see examples \ref{ex10}, \ref{ex104}). 

In this paper we introduce rescaling transformations, which are more general than unitary transformations (which are relevant to isolated systems).
In addition to unitary evolution they also describe irreversible phenomena, which change not only the phase but also the absolute value of the wavefunction (e.g.  quantum tunnelling, the damped/amplified oscillator,  von Neumann projections for quantum measurements, etc). Each of these phenomena has been studied separately as a problem in its own right, for a long time.
Quantum tunnelling has been studied with analytic continuation of a real variable into the complex plane (e.g, \cite{QT1,QT2,QT3,QT4}).
The damped and amplified oscillator  have been studied with various techniques (e.g., with Hamiltonian methods\cite{A1, A2,A3,A4}, with the Langevin equation \cite{B1,B2}, and with other methods\cite{C1,C2,C3}).
Also interaction of a system with the environment (e.g., \cite{D1,D2}) is related to these techniques.
The rescaling transformations are linked to all these phenomena in a unified and abstract way (without the technical details for each of them),
and we show in this paper that they are intimately connected to the Grothendieck bound formalism.

The `Grothendieck philosophy' is to construct abstract mathematical structures, which use the common features and unify many  mathematical areas.
Here we present the Grothendieck bound formalism for a single quantum system, using the rescaling transformations.  
We show that the Grothendieck bound approach provides a general mathematical framework which gives constraints (related to $k_G$) for quantum mechanics 
with rescaling transformations. 
In particular we discuss quantum phenomena which are close to the Grothendieck bound.
The Grothendieck formalism plays complementary role to more specific approaches that use a particular Hamiltonian, or a particular unitary transformation or a particular approach to amplification and damping, etc.

The work involves the following steps:
\begin{itemize}
\item
We introduce rescaling matrices (section \ref{sec31}) which acting on a vector they contract or dilate its length.  Rescaling matrices are suitable for the description of physical phenomena like quantum tunnelling
(e.g., example \ref{ex100}) or damping/amplification (where we have contraction/dilation, e.g., example \ref{ex150}). 
All projectors are rescaling matrices and therefore von Neumann measurements are included in this formalism.
The set of rescaling matrices enlarges significantly the set of unitary matrices, and in addition to unitary transformations (which describe isolated systems) is able to describe a much wider class of physical phenomena 
that includes irreversible phenomena.

\item
We introduce  dequantisation matrices (section \ref{sec32}) which map all vectors in a $d$-dimensional Hilbert space $H(d)$ into vectors in a one-dimensional Hilbert space (which can be viewed as scalars).
In particular superpositions of vectors in $H(d)$ (which are very important in quantum mechanics), are mapped into sums of scalars.
Overall, the dequantisation matrices turn off the Hilbert space formalism and convert it into a formalism of scalars.
\item
In section \ref{sec4}, we express the classical quadratic form ${\cal C}(\theta)$ as the trace of the product of a matrix $\theta$ with two dequantistation matrices (Eq.(\ref{23})), and the quantum' quadratic form ${\cal Q}(\theta)$ 
as the trace of the product of $\theta$ with two rescaling matrices (Eq.(\ref{654})).
Going from ${\cal C}(\theta)$ to ${\cal Q}(\theta)$, replaces the dequantisation matrices with the rescaling matrices, and can be viewed as a passage (Eq.(\ref{450})) from classical physics to quantum physics in its generality, that involves not only unitary transformations in isolated systems, but also quantum tunnelling, damping and amplification, etc.

\item
We introduce the set $G_d$ of matrices $\theta$ that consists of two parts: the subset  $G_d^\prime$ which is invariant under unitary transformations, and the subset $G_d\setminus G_d^\prime$ 
which is {\bf not} invariant under unitary transformations and which is relevant to non-isolated systems interacting with the external world (example \ref{exDC}).
We show (proposition \ref{pro14}) that matrices $\theta\in G_d^\prime$ give ${\cal Q}(\theta)\le 1$, and that only matrices $\theta \in G_d\setminus G_d^\prime$, might give ${\cal Q}(\theta)$ in the ultra-quantum region $(1,k_G)$. 
We explain that ultra-quantumness according to the Grothendieck formalism (i.e.,  ${\cal Q}(\theta)\in (1, k_G)$) is different and complementary to quantumness according to other criteria.

\item

In section \ref{sec2q}, we give an example with ${\cal Q}(\theta)$ in the ultra-quantum region $(1,k_G)$.
It is in the same spirit of examples with ${\cal Q}(\theta)>1$ given in ref.\cite{VV1} in relation to coherent states (which are not discussed here). 
In the present paper the formalism is expanded (in section \ref{sec40}) in order to describe 
phenomena where classically isolated (by high potentials) regions of space, communicate through quantum tunnelling.

\end{itemize}

\section{Notation}  
\begin{enumerate}
\item
Let $H(d)$ be a $d$-dimensional Hilbert space. 
If $v_s$ is a vector in $H(d)$ then its norm is
\begin{eqnarray}
||v_s||=\sqrt{\sum _s|v_s|^2}.
\end{eqnarray}
We use the following notation for the `normalised vector of ones':
\begin{eqnarray}\label{7}
\ket{{\bf J}}=\frac{1}{\sqrt{d}}
\begin{pmatrix}
1\\\vdots\\1
\end{pmatrix}.
\end{eqnarray}
Also
\begin{eqnarray}\label{8}
J_d=d\ket{{\bf J}}\bra{{\bf J}};\;\;\;(J_d)_{rs}=1,
\end{eqnarray}
is the `matrix of ones' (all its elements are equal to one).
\item
If $M$ is a $d\times d$ complex matrix, its Frobenius norm is
\begin{eqnarray}\label{FRO1}
||M||_2=\sqrt{\sum _{r,s}|M_{rs}|^2}.
\end{eqnarray}
It is invariant under unitary transformations.
The following inequality is proved using the Cauchy-Schwartz inequality for the $M_{rs}, K_{rs}$ viewed as $d^2$-dimensional vectors:
\begin{eqnarray}\label{FRO}
|{\rm Tr}(MK)|\le ||M||_2||K||_2.
\end{eqnarray}

\item
Below we will use the inequality
\begin{eqnarray}\label{100}
\left |\sum _{r=1}^da_r\right |^2\le d\sum _{r=1}^d|a_r|^2.
\end{eqnarray}
In the generic case this is a strict inequality, which becomes equality in the special case that the $a_r$ are equal to each other.

This inequality can also be written (for a matrix) as
\begin{eqnarray}\label{111}
\left |\sum _{r,s}a_{rs}\right |^2\le d^2\sum _{r,s}|a_{rs}|^2.
\end{eqnarray}
In the generic case this is a strict inequality, which becomes equality in the special case that the $a_{rs}$ are equal to each other.

\item
We consider a $d$-dimensional Hilbert space $H(d)$ and an orthonormal basis $\ket{X;s}$ which we call position states.
Here $s\in {\mathbb Z}_d$ (the ring of integers modulo $d$), and $X$ is not a variable but it simply indicates position states. 
The Fourier matrix is
\begin{eqnarray}\label{f}
F_{rs}=\frac{1}{\sqrt{d}}\omega^{rs};\;\;\;\omega=\exp\left (\frac{2\pi i}{d}\right );\;\;\;r,s\in{\mathbb Z}_d.
\end{eqnarray}
Acting with the Fourier matrix on the position basis we get another basis which we call momentum basis:
\begin{eqnarray}\label{AA}
\ket{P;r}=\sum _sF_{rs}\ket{X;s}.
\end{eqnarray}
Here $r\in {\mathbb Z}_d$ and $P$ is not a variable but it simply indicates momentum states. 

\item
Let $\pi$ be a permutation 
\begin{eqnarray}
(0,1,...,d-1)\;\overset{\pi} \rightarrow\;(\pi(0),\pi(1),...,\pi(d-1)).
\end{eqnarray}
$\pi$ is an element of the symmetric group ${\cal S}_{d}$ \cite{sagan} which has $d!$ elements.
A permutation matrix $M(\pi)$ is a $d\times d$ matrix with elements 
\begin{eqnarray}\label{4}
[M(\pi)]_{ij}=\delta(i,\pi(j))
\end{eqnarray}
where $\delta$ is the Kronecker delta. 
If $\theta$ is a $d\times d$ matrix, then  
\begin{eqnarray}\label{RRR}
\{[M(\pi)]^\dagger \theta [M(\pi)]\}_{ij}=\theta_{\pi(i),\pi(j)}
\end{eqnarray}
is another matrix whose elements are a permutation of the elements of $\theta$.
The diagonal elements in $\theta$, remain diagonal elements after the transformation.
\end{enumerate}

\section{Rescaling transformations}\label{sec3}
\subsection{Rescaling matrices: enlarging the set of unitary matrices with contractions, dilations and projections}\label{sec31}

Here we enlarge the set of unitary matrices into the rescaling matrices.
\begin{definition}
For any $d\times d$ matrix $V$,
\begin{eqnarray}\label{82}
{\cal N}(V)=\max_i\sqrt{\sum_{j}|V_{ij}|^2}=\max_i\sqrt{(VV^\dagger)_{ii}}
\end{eqnarray}
${\cal S}_d$ is the set of matrices $V$ that have ${\cal N}(V)\le 1$ and which we call `rescaling matrices'.
\end{definition}
\begin{lemma}\label{lem1}
\begin{itemize}
\item[(1)]
If $V,W$ are rescaling matrices then:
\begin{eqnarray}\label{14}
|(VW^\dagger)_{ik}|=|\sum _jV_{ij}W^*_{kj}|\le 1.
\end{eqnarray}

\item[(2)]
If $R,V$ are rescaling matrices, then for any $\lambda\le \frac{1}{\sqrt{d}}$ the $\lambda RV$ is a rescaling matrix.
\end{itemize}
\end{lemma}
\begin{proof}
\begin{itemize}
\item[(1)]
We use the Cauchy-Schwartz inequality for the vectors $V_{ij}$ and $W_{kj}^*$ (with fixed $i,k$).

\item[(2)]
We use Eq.(\ref{100}) and we get
\begin{eqnarray}
\lambda ^2\sum _t|\sum _sV_{rs}R_{st}|^2\le\frac{1}{d}\sum _t|\sum _sV_{rs}R_{st}|^2\le \sum _s\sum _t|V_{rs}R_{st}|^2\le \sum _{s}\left (|V_{rs}|^2\sum _t|R_{st}|^2\right)\le 1.
\end{eqnarray}
\end{itemize}
\end{proof}

We note that:
\begin{itemize}
\item
All $d\times d$ unitary matrices belong in ${\cal S}_d$.
\item
All $d\times d$ projectors $\Pi$ belong in ${\cal S}_d$ (because $(\Pi\Pi^\dagger)_{rr}=\Pi_{rr}\le 1$).
In fact all $\lambda \Pi\in {\cal S}_d$ for all $\lambda\le \lambda_0$ where $\lambda_0=\frac{1}{\max \Pi_{rr}}\ge 1$.

\item
Matrices with  ${\cal N}(V)\le 1$ are rescaling  matrices in the sense that acting on a vector (Eq.(\ref{43}) below) they multiply its length by a factor $0\le \lambda \le \sqrt{d}$ (which is contraction for $0\le \lambda\le 1$ and 
dilation for $1<\lambda\le \sqrt{d}$). Contraction describes physical phenomena like quantum tunnelling or damping where the absolute value of the wave function decreases as a function of some variable. Dilation describes  amplification where the absolute value of the wave function increases as a function of some variable. 
 
\item
If $V$ is an arbitrary matrix, then for any $\lambda\le \frac{1}{{\cal N}(V)}$, the $\lambda V\in {\cal S}_d$.
\item
In the special case $d=1$, ${\cal S}_1$ is the set of complex numbers in the unit disc (in this case the rescaling is contraction).

\end{itemize}
Quantum mechanics in its full generality includes physical phenomena like quantum tunnelling, dissipation and amplification, 
and projections related to measurements.
This requires a wider class of transformations than unitary transformations, which includes contractions and dilations, and projections. The set ${\cal S}_d$ does that, and it is {\bf not} a group because 
it includes irreversible processes.

We can define a `star product' between rescaling matrices $R,V$ as
\begin{eqnarray}
R\star V=\frac{1}{\sqrt{d}}RV\in {\cal S}_d.
\end{eqnarray}
The set ${\cal S}_d$ with the star product is a non-commutative semigroup (without a unity).
Indeed the star product of two rescaling matrices is a rescaling matrix (but the ordinary product might not be a rescaling matrix), and associativity holds.
There is no unit element ($R\star {\bf 1}=\frac{1}{\sqrt{d}}R$).
From a practical point of view, replacing all products with star products will complicate unnecessarily the equations, and we will not use it.

\begin{proposition}
\mbox{}
\begin{itemize}
\item[(1)]
If $\ket{f}$ is a normalised vector in $H(d)$, and $V\in {\cal S}_d$, then
the rescaling matrix $V$ acting on $\bra{f}$ from the right (or $V^\dagger$ acting on $\ket{f}$  from the left)  multiplies its norm by a factor $0\le \lambda \le \sqrt{d}$:
\begin{eqnarray}\label{43}
||V^\dagger\ket{f}||=\lambda;\;\;\;0\le \lambda\le \sqrt{d}.
\end{eqnarray}

\item[(2)]

If $W,V \in {\cal S}_d$, then for any $d\times d$ matrix $\theta$
\begin{eqnarray}\label{44}
||W^\dagger \theta V||_2\le d ||\theta||_2.
\end{eqnarray}

\end{itemize}
\end{proposition}
\begin{proof}
The proofs below use the inequalities in Eqs.(\ref{100}),(\ref{111}).
\begin{itemize}
\item[(1)]
We get
\begin{eqnarray}
\lambda^2=||V^\dagger\ket{f}||^2=\sum _s|\sum _rV_{rs}^*f_r|^2\le d\sum _s\sum _r|V_{rs}^*f_r|^2\le d\sum _{r}\left (|f_r|^2\sum _s|V_{rs}|^2\right)\le d.
\end{eqnarray}
\item[(2)]
We get
\begin{eqnarray}
||W^\dagger\theta V||^2_2&=&\sum _{a,b}|\sum _{r,s}W^*_{ra}\theta_{rs}V_{sb}|^2\le d^2\sum _{a,b}\sum _{r,s}|W_{ra}|^2|\theta_{rs}|^2|V_{sb}|^2\nonumber\\
&=&d^2\sum _{r,s}(\sum_a|W_{ra}|^2)|\theta_{rs}|^2(\sum _b|V_{sb}|^2)\le d^2\sum _{r,s}|\theta_{rs}|^2=d^2||\theta||^2_2.
\end{eqnarray}

\end{itemize}
\end{proof}

\begin{lemma}
\mbox{}
\begin{itemize}
\item[(1)]
If $V$ is a normal matrix then  ${\cal N}(V)={\cal N}(V^\dagger)$, but in general ${\cal N}(V)\ne {\cal N}(V^\dagger)$.
\item[(2)]
If $M(\pi)$ is a permutation matrix then $ {\cal N}(M(\pi)V[M(\pi)]^\dagger)={\cal N}(V)$.
Therefore if $V\in{\cal S}_d$, it follows that $M(\pi)V[M(\pi)]\in {\cal S}_d$.
\item[(3)]
If $U$ is a unitary matrix, then in general ${\cal N}(V)\ne {\cal N}(UVU^\dagger)$.
Therefore if $V\in{\cal S}_d$, the $UVU^\dagger$ might not belong to ${\cal S}_d$.
\end{itemize}
\end{lemma}
\begin{proof}
\mbox{}
\begin{itemize}
\item[(1)]
From Eq.(\ref{82}) follows immediately that if $V$ is a normal matrix, then ${\cal N}(V)={\cal N}(V^\dagger)$. But in general ${\cal N}(V)\ne {\cal N}(V^\dagger)$.
\item[(2)]
${\cal N}(V)=\max_i\sqrt{(VV^\dagger)_{ii}}$ and we use the fact that with permutation transformations the diagonal elements for a matrix remain diagonal (Eq.(\ref{RRR})).
\item[(3)]
If $(VV^\dagger)_{ii}=d_i$, then 
\begin{eqnarray}
{\cal N}(V)=\max_i\sqrt{d_i};\;\;\;{\cal N}(UVU^\dagger)=\max _r\sqrt{\sum_{s,t}U_{rs}(VV^\dagger)_{st}U^*_{rt}}
\end{eqnarray}
The non-diagonal elements of $VV^\dagger$ enter in the calculation of the second expression only, and therefore in general ${\cal N}(V)\ne {\cal N}(UVU^\dagger)$.

\end{itemize}
\end{proof}

\subsection{Dequantisation matrices}\label{sec32}
Let $\{a_i\}$ be a set of $d$ complex numbers in the unit disc. We consider the $d\times d$ matrix
\begin{eqnarray}
{\cal A}_{rs}(a_i)=\frac{1}{\sqrt{d}}a_r;\;\;\;|a_r|\le 1.
\end{eqnarray}
All elements in the same row are equal to each other. The rank of these matrices is one.
We call ${\cal T}_d$ the set of these matrices. It is easily seen that the $|a_r|\le 1$ gives ${\cal N}[{\cal A}(a_i)]\le 1$, and therefore ${\cal T}_d\subset {\cal S}_d$.
We call ${\cal S}_d\setminus {\cal T}_d$ set of `proper' rescaling matrices.

Eq.(\ref{17}) below shows that matrices in ${\cal T}_d$ map all vectors in the Hilbert space $H(d)$ into vectors in one-dimensional vector space (which can be viewed as a set of scalars).
They also map superpositions of vectors (which are at the heart of quantum mechanics), into sums of scalars.
In this sense matrices in ${\cal T}_d$ convert a Hilbert space formalism into a formalism of scalars, and for this reason we call them dequantisation matrices. 

In the same spirit Eq.(\ref{18}) below shows that matrices in ${\cal T}_d$ acting on both sides of an arbitrary matrix, give a scalar times the matrix of ones $J_d$.
So the set of matrices is mapped into a set of scalars.
This is another aspect of the dequantisation.

\begin{lemma}\label{lem2}
Let $ {\cal A}(a_i), {\cal A}(b_i)\in {\cal T}_d$ and $\lambda\le \frac{1}{\sqrt{d}}$. Then: 
\begin{itemize}
\item[(1)]
The product $\lambda {\cal A}(a_i){\cal A}(b_i)\in {\cal T}_d$ 
\item[(2)]
The product $\lambda F{\cal A}(a_i)\in {\cal T}_d$ (where $F$ is the Fourier matrix in Eq.(\ref{f}) which does not belong in ${\cal T}_d$).
\end{itemize}
\end{lemma}
\begin{proof}
\begin{itemize}
\item[(1)]
The $(rs)$-element of $\lambda {\cal A}(a_i){\cal A}(b_i)$ is 
\begin{eqnarray}
\lambda [{\cal A}(a_i){\cal A}(b_i)]_{rs}=\frac{1}{\sqrt d}c_r;\;\;\;c_r=\frac{\lambda}{\sqrt{d}}a_r\sum_t b_t.
\end{eqnarray}
Then
\begin{eqnarray}
|c_r|=\frac{\lambda}{\sqrt{d}}|a_r|\left |\sum_t b_t\right |\le \frac{\lambda}{\sqrt{d}}|a_r|\left (\sum_t |b_t|\right )\le \lambda\sqrt{d} |a_r|\le \lambda\sqrt{d}.
\end{eqnarray}
We want $|c_r|\le 1$ and therefore $\lambda\le \frac{1}{\sqrt{d}}$.
\item[(2)]
The $(rs)$-element of $\lambda F{\cal A}(a_i)$ is 
\begin{eqnarray}
\lambda [F{\cal A}(a_i)]_{rs}=\frac{1}{\sqrt d}c_r;\;\;\;c_r=\frac{\lambda}{\sqrt{d}}\sum_t \omega^{rt}a_t.
\end{eqnarray}
Then
\begin{eqnarray}
|c_r|=\frac{\lambda}{\sqrt{d}}\left |\sum_t \omega ^{rt}a_t\right |\le \frac{\lambda}{\sqrt{d}}\sum_t |a_t|\le \lambda\sqrt{d}.
\end{eqnarray}
We want $|c_r|\le 1$ and therefore $\lambda\le \frac{1}{\sqrt{d}}$.

\end{itemize}
\end{proof}

\begin{proposition}[\bf Dequantisation]
Let $ {\cal A}(a_i), {\cal A}(b_i)\in {\cal T}_d$.
\begin{itemize}
\item[(1)]
If $\ket{f}$ is a normalised vector in $H(d)$, then
\begin{eqnarray}\label{17}
[{\cal A}(a_i)]^\dagger\ket{f}=\lambda\ket{{\bf J}};\;\;\;
\lambda=\sum_ra_r^*f_r;\;\;\;|\lambda|\le \sqrt{d}.
\end{eqnarray}
Here $\ket{{\bf J}}$ is the normalised vector of ones (defined in Eq.(\ref{7})).
Therefore $[{\cal A}(a_i)]^\dagger$ maps the $d$-dimensional space $H(d)$ into the one-dimensional space of vectors $\lambda\ket{{\bf J}}$.
\item[(2)]
For any $d\times d$ matrix $\theta$
\begin{eqnarray}\label{18}
[{\cal A}(a_i^*)]^\dagger \theta {\cal A}(b_i)=\lambda \frac{1}{d}J_d;\;\;\;\lambda=\sum_{r,s} a_r\theta_{rs} b_s.
\end{eqnarray}
Here $J_d$ is the matrix of ones (Eq.({\ref8})). It follows that
\begin{eqnarray}\label{19}
\left |{\rm Tr}\left [[{\cal A}(a_i^*)]^\dagger \theta {\cal A}(b_i)\right ]\right |=\left |\sum_{r,s} a_r\theta_{rs} b_s\right |.
\end{eqnarray}

\end{itemize}
\end{proposition}

\begin{proof}

\begin{itemize}
\item[(1)]
Direct calculation gives the result in Eq.(\ref{17}) with $\lambda=\sum a_r^*f_r$.
Then the Cauchy-Schwartz inequality with the vectors $a_r$ (with $\sum |a_r|^2\le d$) and $f_r$ (with $\sum |f_r|^2=1$) gives $|\lambda|\le \sqrt{d}$.
\item[(2)]
Direct calculation gives the result in Eq.(\ref{18}) and then follows Eq.(\ref{19}).

\end{itemize}
\end{proof}

\subsection{Examples of rescaling matrices}

\begin{example}[\bf tunnelling]\label{ex100}
The purpose of this example is to show how the rescaling matrices introduced in this paper, can be linked to phenomena where classically isolated 
 regions of space (by high potentials), communicate through quantum tunnelling.

The quantum transmission of a particle through a one-dimensional square potential barrier is studied in many quantum mechanics textbooks (e.g.\cite{SC}).
We consider the potential barrier
\begin{eqnarray}\label{31A}
x<0&\rightarrow &V=0\nonumber\\
0<x<a&\rightarrow&V=V_0>0\nonumber\\
x>a&\rightarrow&V=0.
\end{eqnarray}
A particle with mass $m$, momentum $k$ and energy $E=\frac{k^2}{2m}<V_0$ (in units where $\hbar=1$) approaches from the left and is transmitted through the potential (classically it cannot be transmitted because $E<V_0$).
The asymptotic wavefunction is
\begin{eqnarray}\label{RRQ}
x<0&\rightarrow &u_L(x)=A(e^{ikx}+Be^{-ikx})\nonumber\\
x>a&\rightarrow&u_R(x)=ACe^{ikx},
\end{eqnarray}
where $A,AB,AC$ are the amplitudes of the incoming wave, the reflected wave, and the transmitted wave, correspondingly. 
The probability current density is 
\begin{eqnarray}\label{RR}
x<0&\rightarrow &J=\frac{k|A|^2}{m}(1-|B|^2)\nonumber\\
x>a&\rightarrow&J=\frac{k|A|^2}{m}|C|^2;\;\;\;|B|^2+|C|^2=1.
\end{eqnarray}
$J$ is the net flux in each region. Below we impose the normalisation condition 
\begin{eqnarray}
A=\sqrt{\frac{m}{k}}.
\end{eqnarray}

It is known\cite {SC} that
\begin{eqnarray}
&&B=\frac{(k^2-\lambda^2)(1-e^{2i\lambda a})}{(k+\lambda)^2-(k-\lambda)^2e^{2i\lambda a}};\;\;\;\lambda=i\sqrt{2m(V_0-E)}=i\sqrt{2mV_0-k^2}\nonumber\\
&&C=\frac{4k\lambda e^{i(\lambda-k)a}}{(k+\lambda)^2-(k-\lambda)^2e^{2i\lambda a}}.
\end{eqnarray}
We can describe approximately this problem using two `large' but finite dimensional Hilbert spaces $H_L(2j+1)$ and $H_R(2j+1)$ for the left ($x\le 0$) and right ($x\ge a$) regions.
Now the position $x$ and the momentum $k$ take values in ${\mathbb Z}_{2j+1}$.
Using the notation in Eq.(\ref{AA}), we write the result in Eq.(\ref{RRQ})  as
\begin{eqnarray}\label{89A}
x=-2j,...,0&\rightarrow &\ket{u_L}=\sqrt{\frac{m}{k}}(\ket{P;k}+B\ket{P;-k})\nonumber\\
x=a,...,(a+2j)&\rightarrow&\ket{u_R}=\sqrt{\frac{m}{k}}C\ket{P;k}
\end{eqnarray}
From this follows that
\begin{eqnarray}
&&\langle X;x\ket{u_L}=\sqrt{\frac{m}{k}}\frac{1}{\sqrt{2j+1}}(\omega^{xk}+B\omega^{-xk})\nonumber\\
&&\langle X;x\ket{u_R}=\sqrt{\frac{m}{k}}\frac{1}{\sqrt{2j+1}}C\omega^{xk};\;\;\;x,k\in {\mathbb Z}_{2j+1}.
\end{eqnarray}
$\omega$ is defined in Eq.(\ref{f}). They are discrete versions of Eq.(\ref{RRQ}).

In the momentum basis, the state of the incoming wave $\ket{u_L}$ can be written as the column with $2j+1$ elements
\begin{eqnarray}
&&(u_L)_r=0\;{\rm if}\;r\ne k,-k\nonumber\\
&&(u_L)_k=\sqrt{\frac{m}{k}}\nonumber\\
&&(u_L)_{-k}=\sqrt{\frac{m}{k}}B
\end{eqnarray}
and the state of the transmitted wave $\ket{u_R}$ can be written as the column with $2j+1$ elements
\begin{eqnarray}
&&(u_R)_r=0\;{\rm if}\;r\ne k\nonumber\\
&&(u_R)_k=\sqrt{\frac{m}{k}}C.
\end{eqnarray}
Then $\ket{u_R}=W\ket{u_L}$, where $W$ is the $(2j+1)\times (2j+1)$ matrix 
\begin{eqnarray}
&&W_{kk}=C\nonumber\\
&&W_{rs}=0\;{\rm otherwise}.
\end{eqnarray}
Since $|C|\le 1$, this is rescaling non-unitary matrix, and plays similar role to the transmission coefficient.
It is non-invertible matrix as it describes an irreversible phenomenon.

There is no loss of generality if we work in the two-dimensional subspaces $H_L(2)$ of $H_L(2j+1)$ and $H_R(2)$ of $H_R(2j+1)$, that contain quantities with non-zero values.
Then 
\begin{eqnarray}\label{39A}
\ket{u_L}=\sqrt{\frac{m}{k}}\begin{pmatrix}
1\\B
\end{pmatrix};\;\;\;
\ket{u_R}=W\ket{u_L}=\sqrt{\frac{m}{k}}\begin{pmatrix}
C\\0
\end{pmatrix};\;\;\;
W=\begin{pmatrix}
C&0\\
0&0
\end{pmatrix}.
\end{eqnarray}
These vectors are not normalised, and the $\ket{u_L}\bra{u_L}$ and $\ket{u_R}\bra{u_R}$ can be written as $2\times 2$ projectors $\varpi_L, \varpi _R$, times coefficients $\xi_L, \xi_R$ that depend 
on the reflection and transmission parameters $B,C$:
\begin{eqnarray}\label{TT}
\ket{u_L}\bra{u_L}=\xi_L\varpi_L;\;\;\;\xi_L=\frac{m}{k}(1+|B|^2);\;\;\;\varpi _L=\frac{1}{1+|B|^2}
\begin{pmatrix}
1&B^*\\
B&|B|^2
\end{pmatrix},
\end{eqnarray}
and
\begin{eqnarray}\label{TT1}
\ket{u_R}\bra{u_R}=\xi_R\varpi_R;\;\;\;\xi_R=\frac{m}{k}|C|^2;\;\;\;\varpi_R=
\begin{pmatrix}
1&0\\
0&0
\end{pmatrix}.
\end{eqnarray}

We next consider the direct sum $H_L(2)\oplus H_R(2)$ and represent them with the $4\times 4$ matrices (written in a `block notation'):
\begin{eqnarray}
\ket{u_L}\bra{u_L}\rightarrow\xi_L\Pi_L;\;\;\;\Pi_L=
\begin{pmatrix}
\varpi_L&0\\
0&0
\end{pmatrix},
\end{eqnarray}
and 
\begin{eqnarray}
\ket{u_R}\bra{u_R}\rightarrow\xi_R\Pi_R;\;\;\;\Pi_R=
\begin{pmatrix}
0&0\\
0&\varpi_R
\end{pmatrix};\;\;\;\Pi_L\Pi_R=0.
\end{eqnarray}
Then
\begin{eqnarray}\label{TT3}
\xi_L{\cal W}\Pi_L{\cal W}^\dagger=\xi_R\Pi_R ;\;\;\;{\cal W}=\begin{pmatrix}
0&0\\
W&0
\end{pmatrix}.
\end{eqnarray}

We generalise this for the case of many momenta, and represent  quantum mechanics for this problem with $2d$-dimensional Hilbert space $H_L(d)\oplus H_R(d)$, 
where $H_L(d)$ is related to quantum mechanics on the left ($x\le 0$), and $H_R(d)$ is related to quantum mechanics on the right ($x\ge a$).
We do not study in detail the middle region $0\le x\le a$.
The state on the left is described with a matrix $\xi _L\Pi_L$  where $\Pi_L$ is a projector to $H_L(d)$, and $\xi _L$ is a coefficient.
Similarly, the state on the right is described with a matrix $\xi _R\Pi_R$  where $\Pi_R$ is a projector to $H_R(d)$, and $\xi _R$ is a coefficient.
Both $\xi_L, \xi_R$ depend on the transmission and reflection parameters. Then $\Pi_L \Pi_R=0$, and the $\Pi_R$ is related to $\Pi_L$ with a relation analogous to Eq.(\ref{TT3}).
This is discussed from the point of view of the formalism in this paper, in section \ref{sec2q}.

\end{example}

\begin{example}[\bf damping and amplification]\label{ex150}
The purpose of this example is to indicate briefly how the rescaling matrices introduced in this paper (in particular those which are non-unitary matrices), can be linked to the damped and amplified oscillator.

Classically the damped/amplified oscillator (with mass $m=1$) is described with the differential equations
\begin{eqnarray}
\frac{d^2x}{dt^2}+\gamma \frac{dx}{dt}+\omega^2x=0;\;\;\;\frac{d^2y}{dt^2}-\gamma \frac{dy}{dt}+\omega^2y=0
\end{eqnarray}
Here $\omega$ is the frequency of the oscillator in the absence of damping or amplification, and $\gamma\ge 0$ is the damping/amplification  parameter.
The solution of these two equations is
\begin{eqnarray}
x(t)=e^{i\epsilon t}e^{-\frac{1}{2}\gamma t};\;\;\;y(t)=e^{i\epsilon t}e^{\frac{1}{2}\gamma t};\;\;\;\epsilon=\sqrt{\omega ^2-\frac{1}{4}\gamma ^2}.
\end{eqnarray}
We consider the case $\omega>\frac{1}{2}\gamma$ of under-critical damping, so the $\epsilon$ is a real number.

Quantisation considers together the damped and amplified oscillator, and the Hamiltonian $H$ is not Hermitian and has complex eigenvalues with imaginary part related to $\gamma$ (e.g., \cite{A1,A2,A3,A4}).
In this case $\exp(iHt)$ is not unitary.
In the present work we do not need the technical details, but we explain that for the damped/amplified oscillator, the rescaling transformations are non-unitary and the
deviation from unitarity is related to $\gamma$.

\end{example}

\section{The Grothendieck bound formalism in terms of rescaling transformations}\label{sec4}
If $\theta$ is a $d\times d$ complex matrix, the Grothendieck theorem considers the `classical' quadratic form
\begin{eqnarray}\label{89}
{\cal C}(\theta)=\left |\sum_{r,s}\theta _{rs}a_rb_s\right |;\;\;\;|a_r|\le 1;\;\;\;|b_s|\le 1.
\end{eqnarray}
${\cal C}(\theta)$ is a `classical quantity' in the sense that the $a_r, b_s$ are scalars in the unit disc.

The formalism also considers the corresponding `quantum' quadratic form where the scalars are replaced with vectors in the unit ball in a $d$-dimensional Hilbert space $H(d)$:
\begin{eqnarray}\label{34}
{\cal Q}(\theta)=\left |\sum_{r,s}\theta_{rs}\lambda_r\mu_s\bra{u_r}v_s\rangle \right |;\;\;\;\ket{u_r}, \ket{v_r}\in H(d);\;\;\;\lambda_r, \mu_r\le1.
\end{eqnarray}
Here we use the bra-ket notation for normalised vectors, and the $\lambda_r\ket{u_r}$, $\mu_s\ket{v_s}$ are vectors in the unit ball in $H(d)$.
${\cal Q}(\theta)$ is a `quantum quadratic form' in the sense that the scalars have been replaced with vectors.

The Grothendieck theorem states that if the classical ${\cal C}(\theta)\le 1$, then the quantum ${\cal Q}(\theta)$ 
might takes values greater than one, up to the complex Grothendieck constant  $k_G$:
\begin{eqnarray}
{\cal C}(\theta)\le 1\rightarrow {\cal Q}(\theta)\le k_G.
\end{eqnarray}
$k_G$ does not depend on the dimension $d$, and
its exact value is not known. It is known that $k_G\le 1.4049$ and various bounds for its exact value are discussed in \cite{GR5,GR6,GR7}.
$k_G$ is a kind of `ceiling' of the Hilbert space (and quantum) formalism. 

Values of ${\cal Q}(\theta)$ in the region $(1,k_G)$ are of special importance, because this is a classically forbidden region (the ${\cal C}(\theta)$ cannot take values in it).
We refer to the region $(1,k_G)$ as the ultra-quantum region.

Later we rewrite ${\cal C}(\theta)$ and ${\cal Q}(\theta)$ as the trace of $\theta$ times two rescaling matrices.

\subsection{The ultra-quantum set $G_d\setminus G_d^\prime$ related to non-isolated systems}
\begin{definition}
For any $d\times d$ complex matrix $\theta$,
\begin{eqnarray}\label{892}
g(\theta)=\sup \left \{{\cal C}(\theta)=\left |\sum_{r,s}\theta _{rs}a_rb_s\right |;\;\;\;|a_r|\le 1;\;\;\;|b_s|\le 1\right \},
\end{eqnarray}
and
\begin{eqnarray}\label{893}
g^\prime(\theta)=\sup \left \{{\cal C}^\prime (\theta)=\left |\sum_{r,s}\theta _{rs}a_rb_s\right |;\;\;\;\sum_r|a_r|^2\le d;\;\;\;\sum _s|b_s|^2\le d\right \}.
\end{eqnarray}
$G_d$ is the set of $d\times d$ matrices with $g(\theta)\le 1$. $G_d^\prime$ is the set of $d\times d$ matrices with $g^\prime (\theta)\le 1$.
\end{definition}

We note that:
\begin{itemize}
\item
$g^\prime (\theta)= d{\mathfrak s}_{\rm max}$\cite{VV}, where ${\mathfrak s}_{\rm max}$ is the maximum singular value of $\theta$ (for normal matrices 
${\mathfrak s}_{\rm max}=e_{\rm max}$ where $e_{\rm max}$ is the maximum of the absolute values of the eigenvalues of $\theta$).

\item
It is easily seen that
\begin{eqnarray}\label{48}
g(\theta)\le g^\prime (\theta)= d{\mathfrak s}_{\rm max}\Rightarrow G^\prime _d\subseteq G_d.
\end{eqnarray}
The difference $G_d\setminus G_d^\prime$ between these two sets is very important, because we show below 
(proposition \ref{pro14}) that a necessary (but not sufficient) condition for ${\cal Q}(\theta)$ taking values in the ultra-quantum region $(1,k_G)$, is that $\theta\in G_d\setminus G_d^\prime$.
For this reason we call the set $G_d\setminus G_d^\prime$ ultra-quantum.

\item
 If $\theta$ is an arbitrary $d\times d$ complex matrix, then 
\begin{eqnarray}\label{44}
&&\lambda \le \frac{1}{ d{\mathfrak s}_{\rm max}}\;\Rightarrow\;\lambda\theta\in G^\prime_d\nonumber\\
&&\lambda \le \frac{1}{g(\theta)}\;\Rightarrow\;\lambda\theta\in G_d\nonumber\\
&&\lambda > \frac{1}{g(\theta)}\;\Rightarrow\;\lambda\theta\notin G_d.
\end{eqnarray}
If for a matrix $\theta$ the strict inequality $g(\theta)<d{\mathfrak s}_{\rm max}$ holds, then there is a window
\begin{eqnarray}\label{34A}
\frac{1}{ d{\mathfrak s}_{\rm max}}< \lambda< \frac{1}{g(\theta)}\Rightarrow\;\lambda\theta\in G_d\setminus G^\prime _d.
\end{eqnarray}
In isolated systems where we only have unitary transformations, multiplication by a positive number has no physical significance (e.g., the ket and bra vectors always have length equal to one).
But when we consider irreversible phenomena in open (non-isolated) systems, like damping and amplification, multiplication by a number is important. In the present context we see that 
different multiples of the same matrix, can belong to $G_d^\prime$ or to $G_d\setminus G^\prime _d$.

\item
It is very important that the Grothendieck formalism considers the 
\begin{eqnarray}\label{892A}
|a_r|\le 1;\;\;\;|b_s|\le 1,
\end{eqnarray}
in Eqs.(\ref{89}),(\ref{892}), rather than the
\begin{eqnarray}\label{893A}
\sum_r|a_r|^2\le d;\;\;\;\sum _s|b_s|^2\le d.
\end{eqnarray}
 which enters in Eq.(\ref{893}) . Because this enlarges the set of unitary transformations into the rescaling transformations, and incorporates irreversible phenomena like quantum tunnelling, damping and amplification, and projections related to quantum measurements. 
 
 The ultra-quantum set of matrices $G_d\setminus G_d^\prime$ is intimately connected to the difference between Eqs.(\ref{892A}), (\ref{893A}).
$\theta \in G_d\setminus G_d^\prime$ means that $\left |\sum_{r,s}\theta _{rs}a_rb_s\right |$ takes values less or equal to one when the very restrictive Eq(\ref{892A}) holds,
but it might take values greater than one when the less restrictive Eq(\ref{893A}) holds.
\end{itemize}

\begin{lemma}\label{L1}
Let $U$ be a unitary operator. Then:
\begin{itemize}
\item[(1)]
 $g^\prime (\theta)= g^\prime (U\theta U^\dagger)$ and in general $g(\theta)\ne g(U\theta U^\dagger)$. 
\item[(2)]
\begin{itemize}
\item
If $\theta \in G_d^\prime$ then $U\theta U^\dagger\in G_d^\prime$, i.e., 
the set $G_d^\prime$ is  invariant under unitary transformations.
\item
If $\theta \in G_d\setminus G_d^\prime$ then $U\theta U^\dagger$ might not belong to $G_d\setminus G_d^\prime$.
The set $G_d\setminus G_d^\prime$ is not invariant under unitary transformations (physically this indicates that the system is not isolated but it interacts with the external world, and that $\theta$ is some `partial trace'). 
\end{itemize}
\end{itemize}
\end{lemma}
\begin{proof}
\begin{itemize}
\item[(1)]
In Eqs(\ref{892}), (\ref{893}) a unitary transformation $U$ on $\theta$ is equivalent to a transformation on $a_r, b_r$ into
\begin{eqnarray}
A_s=\sum _rU_{sr}^*a_r;\;\;\;B_s=\sum _rU_{sr}b_r.
\end{eqnarray}
The fact that $\sum |a_r|^2\le d$ and $\sum|b_r|^2\le d$ implies that $\sum |A_s|^2\le d$ and $\sum|B_s|^2\le d$.
But the fact that $|a_r|\le 1$ and $|b_r|\le 1$ does {\bf not} imply that $|A_s|\le 1$ and $|B_s|\le 1$.
\item[(2)]
If $\theta \in G_d^\prime$ then by definition $g^\prime (\theta)=1$ and since $g^\prime (\theta)= g^\prime (U\theta U^\dagger)$ it follows that $U\theta U^\dagger\in G_d^\prime$.
This does not hold for $G_d\setminus G_d^\prime$ because $g(\theta)\ne g(U\theta U^\dagger)$. 
\end{itemize}
\end{proof}
We next exemplify our comment above, that $\theta \in G_d\setminus G_d^\prime$ is related to non-isolated systems that interact with the external world.
 We show below that a $2\times 2$ matrix related to Eq.(\ref{TT}) in the tunnelling example \ref{ex100}, belongs to
$G_2\setminus G_2^\prime$ only in the case of non-zero tunnelling.

\begin{example}\label{exDC}
Motivated by Eq.(\ref{TT}) which is related to tunnelling phenomena, we consider the $2\times 2$  matrix 
\begin{eqnarray}
\theta=\frac{m}{k}\begin{pmatrix}
1&B\\
B&B^2
\end{pmatrix};\;\;\;1\ge B>0.
\end{eqnarray}
For simplicity, here $B$ is a real positive number.
We show that the strict inequality $g(\theta)<g^\prime (\theta)$ holds only if $B\ne 1$ (non-zero tunnelling).

$\theta$ has the eigenvalues $e_{\rm min}=0$ and $e_{\rm max}=\frac{m}{k}(1+B^2)$.
Therefore 
$g^\prime (\theta)=\frac{2m}{k}(1+B^2)$.
In order to calculate $g(\theta)$ we take $|a_r|\le 1$ and $|b_r|\le 1$ and we get
\begin{eqnarray}
{\cal C}(\theta)=\frac{m}{k}\left |a_0b_0+Ba_0b_1+Ba_1b_0+B^2a_1b_1\right |\le \frac{m}{k}(1+B)^2.
\end{eqnarray}
This becomes equality when $a_0=a_1=b_0=b_1=1$. Therefore
\begin{eqnarray}
g(\theta)=\frac{m}{k}(1+B)^2.
\end{eqnarray}
Then:
\begin{itemize}
\item
If $B=1$ we have no tunnelling, and $g(\theta)=g^\prime (\theta)=\frac{4m}{k}$. In this case there are no values of $\lambda$ for which $\lambda \theta \in G_2\setminus G_2^\prime$.
\item
If the strict inequality $B<1$ holds we have tunnelling, and the strict inequality $g(\theta)<g^\prime (\theta)$ holds.
In this case $\lambda \theta \in G_2\setminus G_2^\prime$ for
\begin{eqnarray}\label{333}
\frac{k}{2m(1+B^2)}<\lambda\le \frac{k}{m(1+B)^2}.
\end{eqnarray}
\end{itemize}
In this example, the strict inequality $g(\theta)<g^\prime (\theta)$ and $\lambda \theta \in G_2\setminus G_2^\prime$
are intimately related to non-zero tunnelling, and for large tunnelling (small $B$) the window in Eq.(\ref{333}) is large.

\end{example}

\subsection{${\cal C}(\theta)$ and ${\cal Q}(\theta)$ as the trace of a product of matrices}

The Grothendieck theorem can be expressed in terms of the trace of a product of three matrices.
\begin{proposition}
\begin{itemize}
\item[(1)]
${\cal C}(\theta)$ in Eq.(\ref{89}) can be written in terms of dequantisation matrices in ${\cal T}_d$ as
\begin{eqnarray}\label{23}
{\cal C}(\theta)=\left |{\rm Tr}\left [[{\cal A}(a_i^*)]^\dagger \theta {\cal A}(b_i)\right ]\right |.
\end{eqnarray}
\item[(2)]
${\cal Q}(\theta)$ in Eq.(\ref{34}) can be written in terms of matrices $V,W\in {\cal S}_d$ as
\begin{eqnarray}\label{654}
{\cal Q}(\theta)=|{\rm Tr}(\theta VW^\dagger)|.
\end{eqnarray}

\end{itemize}
\end{proposition}
\begin{proof}
\begin{itemize}
\item[(1)]
This follows from Eq.(\ref{19}).
\item[(2)]
The relationship of Eq.(\ref{654}) to Eq.(\ref{34}) is seen if we take $V$ to be a $d\times d$ matrix that has the components of $\mu_s\ket{v_s}$ in the $s$-row
(then $V\in {\cal S}_d$), and $W$ to be a matrix that has the 
components of $\lambda_r\ket{u_r}$ in the $r$-row (then $W\in {\cal S}_d$).
In this case $W^\dagger$ has the complex conjugates of the components of $\lambda_r\ket{u_r}$ in the $r$-column.
Then
\begin{eqnarray}
\lambda_r\mu_s\bra{u_r}v_s\rangle=(VW^\dagger)_{rs}.
\end{eqnarray}

\end{itemize}

\end{proof}

In this formulation, the Grothendieck theorem states that if $\theta\in G_d$ and $V,W\in {\cal S}_d$, then 
\begin{eqnarray}\label{54}
{\cal Q}(\theta)=|{\rm Tr}(\theta VW^\dagger)|\le k_G.
\end{eqnarray}
We note that $\theta\in G_d$ means that $g(\theta)\le 1$, which in turn means that the supremum of all ${\cal C}(\theta)=\left |{\rm Tr}\left [[{\cal A}(a_i^*)]^\dagger \theta {\cal A}(b_i)\right ]\right |$
(for all dequantisation matrices) is less than one.

Arbitrary matrices with appropriate normalisation, belong in $G_d$ and ${\cal S}_d$. We express the Grothendieck theorem  for arbitrary matrices as
\begin{eqnarray}\label{456}
{\cal Q}\left (\frac{\theta}{g(\theta)}\right )=\left |{\rm Tr}\left (\frac{\theta}{g(\theta)}\frac{V}{{\cal N}(V)}\frac{W^\dagger}{{\cal N}(W)}\right )\right |\le k_G.
\end{eqnarray}

The transition from classical to quantum quadratic forms is to go from ${\cal C}(\theta)$ which involves dequantisation matrices in ${\cal T}_d$ (Eq.(\ref{23})),
to $ {\cal Q}(\theta)$ which replaces the dequantisation matrices with general rescaling matrices in ${\cal S}_d$:
\begin{eqnarray}\label{450}
&&{\cal C}(\theta)=\left |{\rm Tr}\left [[{\cal A}(a_i^*)]^\dagger \theta {\cal A}(b_i)\right ]\right |\;\rightarrow\;  {\cal Q}(\theta)=|{\rm Tr}(W^\dagger\theta V)|.
\end{eqnarray}

\begin{proposition}[{\bf Transformations that leave ${\cal Q}(\theta)$ invariant}]\label{pro121}
\mbox{}
${\cal Q}(\theta)$ is {\bf not} invariant under general unitary  transformations $U$:
\begin{eqnarray}
\theta \rightarrow U\theta U^\dagger;\;\;\;V \rightarrow UV U^\dagger;\;\;\;
W^\dagger \rightarrow UW^\dagger U^\dagger.
\end{eqnarray}
But in the special case that $U=M(\pi)$ (Eq.(\ref{4})), ${\cal Q}(\theta)$ is invariant under permutation transformations.
\end{proposition}

\begin{proof}

We use the expression in Eq.(\ref{456}) for ${\cal Q}(\theta)$.
It is easily seen that
\begin{eqnarray}
g(M(\pi)\theta [M(\pi)]^\dagger)= g(\theta);\;\;\;{\cal N}(M(\pi)V [M(\pi)]^\dagger)= {\cal N}(V);\;\;\;{\cal N}(M(\pi)W [M(\pi)]^\dagger)= {\cal N}(W).
\end{eqnarray}
Therefore ${\cal Q}(\theta)$ is invariant under permutations.

For more general unitary transformations
\begin{eqnarray}\label{52}
g(U\theta U^\dagger)\ne g(\theta);\;\;\;{\cal N}(UVU^\dagger)\ne {\cal N}(V);\;\;\;{\cal N}(UWU^\dagger)\ne {\cal N}(W).
\end{eqnarray}
Therefore ${\cal Q}(\theta)$ is {\bf not} invariant under general unitary transformations.
The same argument in terms of the expression in Eq.(\ref{654}) is that if  $\theta \in G_d$ and $V,W\in {\cal S}_d$,  the transformed matrices after a unitary transformation might not belong 
to $G_d$ and ${\cal S}_d$.

\end{proof}

\subsection{Grothendieck ultra-quantumness is different from quantumness according to other criteria}

From the Grothendieck formalism point of view, `ultra-quantumness' is related to ${\cal Q}(\theta)\in (1,k_G)$ (because ${\cal C}(\theta)$ cannot take values in it).
Here we show that important physical examples which involve isolated systems and show quantum behaviour according to a different criterion for quantumness, give ${\cal Q}(\theta)\le 1$.

\begin{example}\label{ex10}
Let $\theta=\ket{f}\bra{g}$ which in some basis is written as the matrix
\begin{eqnarray}
\theta_{rs}=f_rg_s^*;\;\;\;\sum _r|f_r|^2=\sum _r|g_r|^2=1.
\end{eqnarray}
Then for $V=U$ where $U$ is a unitary matrix (in which case $V\in {\cal S}_d$) and $W={\bf 1}\in{\cal S}_d$ we prove that
\begin{eqnarray}\label{123}
{\cal Q}\left (\frac{\theta}{g(\theta)}\right )=\left |\frac{{\rm Tr}(\theta U)}{g(\theta)}\right |\le 1.
\end{eqnarray}
It is easily seen that ${\rm Tr}(\theta U)\le 1$. We next show that $g(\theta)>1$. Indeed 
\begin{eqnarray}
{\cal C}(\theta)=\left |\sum_{r,s}f_rg_s^* a_rb_s\right |;\;\;\;|a_r|\le 1;\;\;\;|b_s|\le 1.
\end{eqnarray}
We choose $a_r, b_s$ such that $f_ra_r=|f_r|$ and $g_s^*b_s=|g_s|$.
Then 
\begin{eqnarray}\label{PO}
g(\theta)\ge {\cal C}(\theta)=\sum_r|f_r|\sum _s|g_s|\ge 1.
\end{eqnarray}
This proves Eq.(\ref{123}).

For some unitary operators  ${\rm Tr}(\theta U)$ shows quantum interference in the form of strongly oscillatory behaviour (e.g., \cite{OS1,OS2}).
This is an important quantum feature, and yet seen from the Grothendieck point of view it gives ${\cal Q}\le 1$.
 In this sense `quantumness' associated with values of ${\cal Q}$ near the Grothendieck bound, is different from `quantumness' associated to quantum interference.
\end{example}

\begin{example}\label{ex104}
Fourier transform underpins Harmonic Analysis and Quantum Mechanics which is based on it.
One aspect of this is the uncertainty relations. For a density matrix $\theta$ with small uncertainty (quantified in various ways),  the uncertainty for its Fourier dual density matrix
${\widetilde \theta}=F\theta F^\dagger$ is large (here $F$ is the Fourier transform in Eq.(\ref{f})).
This general idea is called `squeezing'.

Motivated by this we explored whether for a density matrix $\theta$ with ${\cal Q}(\theta)\le 1$, 
its Fourier dual density matrix $\widetilde \theta$ has ${\cal Q}(\widetilde \theta)\ge 1$.
From lemma \ref{L1} follows that if $\theta \in G_d^\prime$ then ${\widetilde \theta}\in G_d^\prime$.
Then according to proposition \ref{pro14} below, both ${\cal Q}(\theta)\le 1$ and ${\cal Q}({\widetilde \theta})\le 1$.
Fourier duality does not play an important role for `ultra-quantumness' (which is ${\cal Q}(\theta)\in (1, k_G)$).
Grothendieck ultra-quantuness is different from quantumness according to
uncertainty relations and squeezing.
\end{example}

\begin{example}
This is a special case of example \ref{ex10}.
We show that we can have values of ${\cal Q}$ smaller than the corresponding values of ${\cal C}$.

We take $f_r=g_s=\frac{1}{\sqrt{d}}$and then 
\begin{eqnarray}
\theta=\frac{1}{d}J_d.
\end{eqnarray}
Here $J_d$ is the `matrix of ones'.
In this case
\begin{eqnarray}
{\cal C}(\theta)=\frac{1}{d}\left |\sum_{r,s}a_rb_s\right |.
\end{eqnarray}
We choose $a_r=b_s=1$ and we get ${\cal C}(\theta)=d$. Therefore $g(\theta)\ge d$.

Also
\begin{eqnarray}
\left |{\rm Tr}(\theta U)\right |=\frac{1}{d}\left |\sum_{r,s}U_{rs}\right |\le \frac{1}{d}\sum_{r,s}|U_{rs}|
\end{eqnarray}
Since $U$ is a unitary matrix
\begin{eqnarray}
\sum_s|U_{rs}|^2=1\;\rightarrow\;\sum_s|U_{rs}|\le \sqrt{d}\;\rightarrow\;\frac{1}{d}\sum_{r,s}|U_{rs}|\le \sqrt{d}.
\end{eqnarray}
Therefore $\left |{\rm Tr}(\theta U)\right |\le \sqrt{d}$ and 
\begin{eqnarray}
{\cal Q}\left (\frac{\theta}{g(\theta)}\right )=\left |\frac{{\rm Tr}(\theta U)}{g(\theta)}\right |\le \frac{1}{\sqrt{d}}.
\end{eqnarray}
In this example ${\cal C}\left (\frac{\theta}{g(\theta)}\right )\in(0,1)$ and ${\cal Q}\left (\frac{\theta}{g(\theta)}\right )\in(0,\frac{1}{\sqrt d})$.

\end{example}

\begin{example}\label{ex14}
This example is needed in the proof of proposition \ref{pro14} below.

Let $\theta$ be the diagonal matrix 
\begin{eqnarray}
\theta={\rm diag}(z_0,...,z_{d-1});\;\;\;\sum _r|z_r|=1;\;\;\;z_r\in {\mathbb C};\;\;\;r\in{\mathbb Z}_d.
\end{eqnarray}
All diagonal density matrices are special cases. Then $g(\theta)=1$ (therefore $\theta\in G_d$) and 
${\cal Q}(\theta)\le 1$.

We first prove that $g(\theta)=1$.  Indeed ${\cal C}(\theta)\le 1$:
\begin{eqnarray}
{\cal C}(\theta)=\left |\sum_{r} z_ra_rb_r\right |\le \sum_{r} |z_r|\cdot |a_rb_r|\le 1;\;\;\;|a_r|\le 1;\;\;\;|b_s|\le 1.
\end{eqnarray}
If $z_r=|z_r|\exp(i\phi_r)$, we chose $a_rb_r=\exp(-i\phi_r)$ and we get ${\cal C}(\theta)=1$. Therefore $g(\theta)=1$, and $\theta\in G_d$.

Then
\begin{eqnarray}
&&{\cal Q}(\theta)=\left |\sum_{r}z_r\lambda_r\mu_r\bra{u_r}v_r\rangle \right |\le \sum_{r}|z_r|\lambda_r\mu_r|\bra{u_r}v_r\rangle |\le 1\nonumber\\
&&\ket{u_r}, \ket{v_r}\in H(d);\;\;\;\lambda_r, \mu_r\le1.
\end{eqnarray}
If $z_r=|z_r|\exp(i\phi_r)$, we chose $\ket{v_r}=\exp(-i\phi _r)\ket{u_r}$ and $\lambda_r=\mu_r=1$, and we get the maximum value one.
This proves the statement.

\end{example}

\section{A necessary condition for ${\cal Q}(\theta)> 1$}\label{secNE}

We start by showing that for matrices $\theta\in G_d^\prime$ and any $V,W\in {\cal S}_d$, we get ${\cal Q}(\theta)\le 1$.
Therefore a necessary (but not sufficient) condition for ${\cal Q}(\theta)> 1$ is that $\theta \in G_d\setminus G_d^\prime$.
This has been proved in ref\cite{VV} for the special case $W={\bf 1}$ (because this involves ${\rm Tr}(\theta V)$ which describes many physical quantities).
Here it is proved in the general case, and complemented with extra conditions.

\begin{proposition}\label{pro14}
Let $\theta$ be  a normal matrix (in which case $g(\theta)\le de_{\rm max}$) and $V,W\in {\cal S}_d$.
\begin{itemize}
\item[(a)]
 If $\lambda\le \frac{1}{de_{\rm max}}$ in which case $\lambda\theta\in G_d^\prime$, we get
\begin{eqnarray}\label{430}
{\cal Q}(\lambda \theta)=|{\rm Tr}(\lambda\theta VW^\dagger)|\le 1.
\end{eqnarray}
\item[(b)]
 If $\lambda\le \frac{1}{g(\theta)}$ in which case $\lambda\theta\in G_d$, a necessary (but not sufficient) condition for  
\begin{eqnarray}
{\cal Q}(\lambda \theta)=|{\rm Tr}(\lambda\theta VW^\dagger)|>1
\end{eqnarray}
is the following requirements:
\begin{enumerate}
\item
 That  the strict inequality $g(\theta)<de_{\rm max}$ holds and that $\lambda$ takes values in the interval
\begin{eqnarray}\label{101}
\frac{1}{de_{\rm max}}<\lambda<\frac{1}{g(\theta)},
\end{eqnarray}
so that $\lambda\theta\in G_d\setminus G_d^\prime$.
\item
That $\lambda\theta$ has non-zero off-diagonal elements (of course $\lambda\theta$ can be diagonalised but as explained in lemma \ref{L1} this would take it outside the set $G_d$ and then Eq.(\ref{54}) is not applicable).
\item
That at least one of the $V,W$ is a proper rescaling matrix (it belongs to ${\cal S}_d\setminus {\cal T}_d$).
\end{enumerate}
\end{itemize}
\end{proposition}
\begin{proof}
\begin{itemize}
\item[(a)]
We regard the $\theta VW^\dagger$ as the product of $\theta V$ with $W^\dagger$ and using Eq.(\ref{FRO}) we get
\begin{eqnarray}
|{\rm Tr}(\theta VW^\dagger)|\le ||\theta V||_2||W^\dagger||_2.
\end{eqnarray}
Since $W\in{\cal S}_d$, it follows that $||W^\dagger||_2\le \sqrt{d}$.
We next diagonalise the matrix $\theta$ as
$\theta =U{\widehat \theta}U^\dagger$,
where $\widehat \theta$ is a diagonal matrix, and $U$ is a unitary matrix.
Since the Frobenius norm is invariant under unitary transformations, we get
\begin{eqnarray}
 ||\theta V||_2=||\widehat \theta\widehat V||_2=\sqrt{\sum _{r,s}|\widehat \theta_{rr}|^2|\widehat V_{rs}|^2};\;\;\;\widehat V=UVU^\dagger.
\end{eqnarray}
Since $|\widehat \theta_{rr}|\le e_{\rm max}$ where $e_{\rm max}$ is the maximum of the absolute values of the eigenvalues of $\theta$, we get
\begin{eqnarray}
 ||\theta V||_2\le e_{\rm max}\sqrt{\sum _{r,s}|\widehat V_{rs}|^2}=e_{\rm max}||\widehat V||_2=e_{\rm max}||V||_2\le \sqrt{d}e_{\rm max}. 
 \end{eqnarray}
We used here the fact that  $V\in{\cal S}_d$, and therefore $||V||_2\le \sqrt{d}$. It follows that
\begin{eqnarray}
 \frac{1}{de_{\rm max}}|{\rm Tr}(\theta VW^\dagger)|<1.
\end{eqnarray}
\item[(b)]
\begin{enumerate}
\item
We consider the case that $\lambda\theta\in G_d$.
From the first part, follows immediately that a necessary (but not sufficient) condition for ${\cal Q}(\theta)=|{\rm Tr}(\lambda\theta V)|>1$ is that $\lambda \theta \in G_d\setminus G_d^\prime$.
For this it is necessary that the strict inequality $g(\theta)<de_{\rm max}$ holds, so that there is a window of values of $\lambda$ (those in Eq.(\ref{34A})) for which $\lambda \theta \in G_d\setminus G_d^\prime$.
\item
We have seen in example \ref{ex14} that diagonal $\theta$ would give ${\cal Q}(\theta)\le 1$.
Therefore $\theta$ should have non-zero off-diagonal elements.
We note that since $\lambda \theta \in G_d\setminus G_d^\prime$, diagonalisation would take $\lambda\theta$ outside the set $G_d$ (see lemma \ref{L1}).
\item

If both $V,W\in {\cal T}_d$ then ${\cal Q}(\lambda \theta)={\cal C}(\lambda\theta)\le 1$.
Therefore at least one of the $V,W$ should belong to ${\cal S}_d\setminus {\cal T}_d$.
\end{enumerate}
\end{itemize}
\end{proof}

The following example shows that the condition in proposition \ref{pro14} is {\bf not} sufficient.

\begin{example}
In example \ref{exDC} we have seen that $\lambda\theta\in G_2\setminus G_2^\prime$ when $\lambda$ in the interval of Eq(\ref{333}).
We show that for any $V,W\in {\cal S}_2$ the ${\cal Q}(\lambda \theta)\le 1$.
We consider the following general matrices in ${\cal S}_2$
\begin{eqnarray}
V=\begin{pmatrix}
a_1&b_1\\
c_1&d_1
\end{pmatrix};\;\;|a_1|^2+|b_1|^2\le 1;\;\;\;|c_1|^2+|d_1|^2\le 1,
\end{eqnarray}
and 
\begin{eqnarray}
W=\begin{pmatrix}
a_2&b_2\\
c_2&d_2
\end{pmatrix};\;\;|a_2|^2+|b_2|^2\le 1;\;\;\;|c_2|^2+|d_2|^2\le 1.
\end{eqnarray}
We take the maximum value of $\lambda$ within the interval in Eq.(\ref{333}) and then
\begin{eqnarray}
\lambda \theta=\frac{1}{(1+B)^2}\begin{pmatrix}
1&B\\
B&B^2
\end{pmatrix};\;\;\;1> B>0.
\end{eqnarray}
Therefore
\begin{eqnarray}
{\rm Tr}(\lambda \theta VW^\dagger)=\frac{1}{(1+B)^2}
[(a_1a_2^*+b_1b_2^*)+B(a_1c_2^*+c_1a_2^*+b_1d_2^*+d_1b_2^*)+B^2(c_1c_2^*+d_1d_2^*)]
\end{eqnarray}
It follows that
\begin{eqnarray}
|{\rm Tr}(\lambda \theta VW^\dagger)|\le \frac{1}{(A+B)^2}
[|a_1a_2^*+b_1b_2^*|+B|a_1c_2^*+c_1a_2^*|+B|b_1d_2^*+d_1b_2^*|+B^2|c_1c_2^*+d_1d_2^*|]
\end{eqnarray}
The Cauchy-Schwartz inequality shows that
\begin{eqnarray}
|a_1a_2^*+b_1b_2^*|\le1;\;\;\;|a_1c_2^*+c_1a_2^*|\le 1;\;\;\;|b_1d_2^*+d_1b_2^*|\le 1;\;\;\;|c_1c_2^*+d_1d_2^*|\le 1.
\end{eqnarray}
Therefore $|{\rm Tr}(\lambda \theta VW^\dagger)|\le 1$ for all values of $\lambda$ within the interval in Eq.(\ref{333}).
It is seen that allthough $\lambda\theta\in G_2\setminus G_2^\prime$, the ${\cal Q}(\lambda \theta)\le 1$ for all matrices in ${\cal S}_2$.

\end{example}

\section{Example with ${\cal Q}(\theta)>1$ related to tunnelling}\label{sec2q}

The example in this section gives ${\cal Q}(\theta)\in (1, k_G)$, and has been presented in ref.\cite{VV1} in relation to the coherent states discussed there (which are not used here).
The formalism is expanded here in a different direction (in section \ref{sec40} below), so that it can be used for the description of tunnelling phenomena that generalise example \ref{ex100}.
It is shown that  ultra-quantum phenomena according to the Grothendieck bound, include phenomena where two parts of the space which are classically isolated by high potentials, communicate through tunnelling.

We consider the matrix
\begin{eqnarray}\label{AQ1}
M(z)=\frac{1}{2}
\setcounter{MaxMatrixCols}{12}
\begin{pmatrix}
1&z&0&1&-z&0\\
z&0&1&-z&0&1\\
0&1&z&0&1&-z
\end{pmatrix};\;\;\;M(z)[M(z)]^\dagger={\bf 1}_3;\;\;\;|z|=1.
\end{eqnarray}
Here $z$ is a constant with $|z|=1$.
Also
\begin{eqnarray}
[M(z)]^\dagger M(z)=\Pi(z)=\frac{1}{4}
\begin{pmatrix}
2&z&z^*&0&-z&z^*\\
z^*&2&z&z^*&0&-z\\
z&z^*&2&-z&z^*&0\\
0&z&-z^*&2&-z&-z^*\\
-z^*&0&z&-z^*&2&-z\\
z&-z^*&0&-z&-z^*&2
\end{pmatrix},
\end{eqnarray}
is a $6\times 6$ projector with eigenvalues $1$ (with multiplicity $3$) and $0$ (with multiplicity $3$).
We call $M(z)$ a semi-unitary matrix in the sense that $M(z)[M(z)]^\dagger$ is a projector (which can be viewed as unit matrix within the space it projects into) and $[M(z)]^\dagger M(z)={\bf 1}_3$.

The projector $\Pi(z)$ acts on a six-dimensional space $H(6)$.
The eigenvectors of $\Pi(z)$ corresponding to the eigenvalue one (zero) span a $3$-dimensional space $H(3)$ ($H(3)_{\rm null}$), and
\begin{eqnarray}
H(6)=H(3)\oplus H(3)_{\rm null}.
\end{eqnarray}

\subsection{Application to tunnelling phenomena}\label{sec40}

We first prove (with direct multiplication and addition) the following relations which are important here:
\begin{eqnarray}
\Pi(z)+\Pi(-z)={\bf 1}_6;\;\;\;\Pi(-z)\cdot\Pi(z)=0;\;\;\;M(-z)[M(z)]^\dagger=0;\;\;\;|z|=1.
\end{eqnarray}
Therefore ${\bf 1}_6-\Pi(z)=\Pi(-z)$ is the projector to $H(3)_{\rm null}$.

We now use the above formalism for  tunnelling problems which generalise  example \ref{ex100}.
We use the space $H(3)$ and the projector $\xi_L\Pi(z)$  (which is related to the matrix $M(z)$) to describe quantum mechanics on the left ($x<0$) of the potential barrier in Eq.(\ref{31A}).
We also use the space $H(3)_{\rm null}$ and the projector $\xi_R[{\bf 1}_6-\Pi(z)]=\xi_R\Pi(-z)$  (which is related to the matrix $M(-z)$) to describe quantum mechanics on the  right ($x>a$) of the potential barrier.
The $\xi_L, \xi_R$ depend on the reflection and transmission parameters (for the simple example \ref{ex100}, see Eqs.(\ref{TT}),(\ref{TT1})).

We use Eq.(\ref{654}) with
\begin{eqnarray}
\theta_L=\xi_L\Pi(z);\;\;\;V=W=\Pi(z)\sqrt{2}\in {\cal S}_6.
\end{eqnarray}
It is easily seen that ${\cal N}[\Pi(z)\sqrt{2}]=1$ and therefore $\Pi(z)\sqrt{2}\in {\cal S}_6$.
Using Eq.(\ref{48}) we find $g^\prime [\Pi(z)]=6$. It has been proven in \cite{VV1}
 that for {\bf generic} $z$ the strict inequality $g[\Pi(z)]<6$ holds. It follows that for
\begin{eqnarray}\label{TTT}
\frac{1}{6}<\xi_L\le \frac{1}{g[\Pi(z)]},
\end{eqnarray}
the $\xi_L\Pi(z)\in G_6\setminus G_6^\prime$. If we have an explicit expression of the $\xi_L$ as a function of the reflection and transmission parameters (generalisations of Eqs.(\ref{TT}),(\ref{TT1})), 
Eq.(\ref{TTT}) will lead to intervals for these parameters.

We now get
\begin{eqnarray}
{\cal Q}(\theta_L)=|{\rm Tr}[\theta_L(\Pi(z)\sqrt{2})(\Pi(z)\sqrt{2})]|=6\xi_L,
\end{eqnarray}
and in the interval of Eq.(\ref{TTT}) we get ${\cal Q}(\theta_L)>1$.
So there are matrices in ${\cal S}_6$ which give ${\cal Q}(\theta_L)>1$, and this indicates that the system interacts with an external system.

The above formalism holds for any $z$, and therefore for $-z$ which as shown above is linked to the region on the right of the potential.
Then the analogue of Eq.(\ref{TT3}) is
\begin{eqnarray}
\Pi(z)=W^\dagger\Pi(-z)W;\;\;\;W=[M(-z)]^\dagger M(z).
\end{eqnarray}

\section{Discussion}

We linked the Grothedieck bound formalism for a single quantum system, to `rescaling transformations'. They enlarge the set of unitary transformations, with transformations which 
are related to irreversible phenomena, like quantum tunnelling, damping and amplification, etc. 
Examples have been given of rescaling matrices which are linked to quantum tunnelling (example \ref{ex100}) and damping and amplification (example \ref{ex150}).

A special case of the rescaling matrices are the dequantisation matrices (section \ref{sec32}).
They map the Hilbert space formalism to a formalism of scalars.

The quantum quadratic form ${\cal Q}(\theta)$ has been expressed as the trace of the product of $\theta$ with two rescaling matrices (Eq.(\ref{654})).
Also the classical quadratic form ${\cal C}(\theta)$ has been expressed as the trace of the product of $\theta$ with two dequantisation matrices (Eq.(\ref{23})). 
In this approach semiclassical limit is to replace the rescaling matrices with dequantisation matrices (Eq.(\ref{450})).

When ${\cal C}(\theta)\le 1$, the ${\cal Q}(\theta)$ can take values greater than one up to the Grothendieck constant $k_G$.
Examples with ${\cal Q}(\theta)$ in the ultra-quantum region  $(1,k_G)$ are very important, because this  region is classically forbidden (${\cal C}(\theta)$ cannot take values in it).
A necessary (but not sufficient) condition for ${\cal Q}(\theta)>1$ has been given in section \ref{secNE}. 

An example with ${\cal Q}(\theta)\in (1,k_G)$ is given in section \ref{sec2q}.
Similar examples in larger Hilbert spaces (with ${\cal Q}(\theta)>1$) have been given in ref.\cite{VV1}, in relation to coherent states. 
In the present paper the formalism has been expanded (in section \ref{sec40}) in order to describe
phenomena where classically isolated (by high potentials) regions of space, communicate through quantum tunnelling.

In this paper the abstract mathematical Grothendieck formalism, has been expressed in terms of rescaling and dequantisation transformations, 
and linked to a large variety of physical phenomena (including irreversible phenomena) without the technical details for each of them.

\end{document}